\documentstyle[prl,aps,epsfig]{revtex}
\begin{document}
\draft
\author{Mikio Eto$^{1,2}$ and Yuli V.\ Nazarov$^{1}$}
\address{$^1$Department of Applied Physics/DIMES,
Delft University of Technology, \\
Lorentzweg 1, 2628 CJ Delft, The Netherlands \\
$^2$Faculty of Science and Technology, Keio
University, \\
3-14-1 Hiyoshi, Kohoku-ku, Yokohama 223-8522, Japan}
\title{Enhancement of Kondo Effect in Quantum Dots with 
an Even Number of Electrons}
\date{\today}
\maketitle
\begin{abstract}
We investigate the Kondo effect in a quantum dot with almost
degenerate spin-singlet and triplet states for an even number of electrons.
We show that the Kondo temperature as a function of
the energy difference between
the states $\Delta$ reaches its maximum around $\Delta=0$
and decreases with increasing $\Delta$.
The Kondo effect is thus enhanced by competition between singlet and
triplet states.
Our results explain recent experimental findings.
We evaluate the linear conductance in the perturbative regime.
\end{abstract}
\pacs{73.23.Hk, 72.15.Qm, 85.30.Vw}

The Kondo effect \cite{classics} takes place when a localized spin $S$
is brought in contact with electron Fermi sea. The Kondo effect 
gives rise to a new many-body ground
state that has a lesser spin. Recently
the Kondo effect has been observed in semiconductor quantum dots connected
to external leads by tunnel junctions \cite{Kondo1}.
In this case the localized spin is formed by electrons in the dot.
The number of electrons $N$ is fixed by Coulomb blockade to integer values
and can be tuned by gate voltage. 
Despite the Coulomb blockade, 
the ground state of the dot
is usually similar to what one obtains disregarding the interaction.
The discrete spin-degenerate levels in the dot are 
consecutively occupied, and the total spin is zero or 1/2
for an even and odd number of electrons, respectively.
Then the Kondo effect takes place only in the latter case
\cite{Glazman,Ng}.

Significant deviations from this plain picture 
were recently observed in so-called
``vertical'' quantum dots \cite{Tarucha,Leo}.
The strength of the electron-electron Coulomb
interaction in such dots is comparable with 
the spacing of discrete levels, and this may give rise to a complicated
ground state.
For example, if two electrons are put into nearly degenerate levels, 
the exchange interaction favors a spin-triplet state. 
This state can be changed to a 
spin singlet by applying a magnetic field since the magnetic field
increases the level spacing \cite{Tarucha}.

This gives a unique possibility to change the spin of the ground state
during an experiment and even obtain extra degenerate states
by tuning the energies of different spin configurations to the same value. 
Such a possibility hardly exists 
in the traditional solid state context.
The Kondo effect in multilevel quantum dots has been investigated
by several groups \cite{Inoshita,Pustilnik}.
In this letter we examine a novel effect that stems from the competition
between spin-singlet and threefold spin-triplet states
for an even number of electrons in a dot.
Since the energy difference
between the states $\Delta$ can be
controlled experimentally, 
we elucidate $\Delta$ dependence of Kondo
temperature as a typical energy scale for the Kondo effect
and linear conductance through the dot.
This enables direct comparison between our calculations and
recent experimental results \cite{exp}.

At large positive $\Delta$ the dot is in a triplet ground state
and an extra singlet state can be disregarded. The Kondo effect
follows a usual $S=1$ scenario.
At large negative $\Delta$ the ground state of the dot is 
a spin singlet and the Kondo effect ceases to exist. 
From this one could suggest that the Kondo temperature decreases as 
$\Delta \rightarrow 0$ from the positive side. 
Our results
show just the opposite. The Kondo temperature $T_K(\Delta)$ is enhanced
at small $\Delta$ and reaches its maximum at $\Delta \approx T_K^{max}$.
At $\Delta \gg T_K^{max}$ the Kondo temperature decreases with increasing
$\Delta$ obeying a power law $T_K(\Delta) \propto 1/\Delta^{\gamma}$. 
The exponent $\gamma$ is not universal but depends on model parameters.
Our results clearly demonstrate the importance of one of the basic
principles of Kondo physics: although the Kondo effect occurs at small
energy scale $T_K$, the value of this scale is determined by 
all energies from $T_K$ up to the upper cutoff. In our case, the energies
from $\Delta$ to the upper cutoff would feel 4-fold degeneracy of the dot
states, which enhances the Kondo temperature.
At $\Delta < \Delta_c$ ($|\Delta_c| \sim T_K^{max}$),
the Kondo effect is not relevant.
We stress the difference between this mechanism and one
proposed in Ref.\ \cite{Pustilnik}, where the Kondo effect arises from extra
degeneracy between one component of the spin-triplet state and
a singlet state, which is brought by the Zeeman splitting.

To model the situation, it is sufficient to consider two extra electrons
in a quantum dot at the background of a singlet state of all other
$N-2$ electrons, which we will regard as the vacuum. 
These two extra electrons occupy two levels of different orbital
symmetry  
\cite{symmetry}.
The energies of the levels are $\varepsilon_1, \varepsilon_2$.
Possible two-electron states are (i) the threefold spin-triplet state, 
(ii) the spin-singlet state of the same orbital symmetry as the triplet state,
$1/\sqrt{2}(d_{1 \uparrow}^{\dagger} d_{2 \downarrow}^{\dagger}
 -d_{1 \downarrow}^{\dagger} d_{2 \uparrow}^{\dagger}) |0 \rangle$,
and (iii) two singlets of different orbital symmetry,
$d_{1 \uparrow}^{\dagger} d_{1 \downarrow}^{\dagger} |0 \rangle$,
 $d_{2 \uparrow}^{\dagger} d_{2 \downarrow}^{\dagger} |0 \rangle$.
Among the singlet states,
we only consider a state of the lowest energy, which belongs 
to the group (iii). Thus we restrict our attention to four 
states, $|S M \rangle$: 
\begin{eqnarray}
|1 1 \rangle & = & d_{1 \uparrow}^{\dagger} d_{2 \uparrow}^{\dagger}
|0 \rangle  \\
|1 0 \rangle & = & \frac{1}{\sqrt{2}} (d_{1 \uparrow}^{\dagger} d_{2
\downarrow}^{\dagger}
                        +d_{1 \downarrow}^{\dagger} d_{2
\uparrow}^{\dagger}) |0 \rangle \\
|1 -1 \rangle & = & d_{1 \downarrow}^{\dagger} d_{2 \downarrow}^{\dagger}
|0 \rangle \\
|0 0 \rangle & = & \frac{1}{\sqrt{2}} (C_1 d_{1 \uparrow}^{\dagger}
d_{1 \downarrow}^{\dagger}
 -C_2 d_{2 \uparrow}^{\dagger} d_{2 \downarrow}^{\dagger}) |0 \rangle,
\label{eq:singlet}
\end{eqnarray}
where $d_{i \sigma}^{\dagger}$ creates an electron with spin $\sigma$ in
level $i$.
Their energies, $E_{S=1}$, $E_{S=0}$,
and coefficients in the singlet state,
$C_1$, $C_2$ ($|C_1|^2+|C_2|^2=2$),
are determined by the electron-electron interaction
and one-electron level spacing $\delta=\varepsilon_2-\varepsilon_1$.
At the moment, we set $C_1=C_2=1$ \cite{com1} and show that this is the
general case afterwards.
The energy difference, $\Delta=E_{S=0}-E_{S=1}$, is changed by applying
a magnetic field $B$ as shown in Fig.\ 1(a). 
We disregard the Zeeman splitting of spin states
since this is a much smaller energy scale than the orbital effect of the
magnetic field in semiconductor heterostructures in use \cite{Tarucha,Leo}.
The exact condition for this is $E_{\rm{Zeeman}} \ll T_K$. 

The dot is connected to two external leads $L$, $R$ with free electrons
being described by
\begin{equation}
H_{\rm{leads}}=\sum_{\alpha=L,R}\sum_{k \sigma i} \varepsilon_{\alpha k}^{(i)}
c_{\alpha, k\sigma}^{(i) \dagger}
c_{\alpha, k\sigma}^{(i)}.
\end{equation}
The tunneling between the dot and the leads is written as
\begin{equation}
H_T=\sum_{\alpha=L,R} \sum_{k \sigma i} (V_{\alpha,i}c_{\alpha,
k\sigma}^{(i) \dagger} d_{i \sigma} + \rm{H.c.}).
\end{equation}
Here $c_{\alpha, k\sigma}^{(i) \dagger}$ is the creation operator of an
electron in lead $\alpha$ with
momentum $k$, spin $\sigma$, and orbital symmetry $i$ $(=1,2)$. 
We assume that the orbital symmetry is conserved
in the tunneling processes. Therefore we have two electron ``channels''
in each lead.

We assume that the state of the dot with $N$ electrons is stable, so that
addition/extraction energies,
$E^{\pm} \equiv E(N \pm 1) -E(N) \mp \mu$
where $\mu$ is the Fermi energy in the leads,
are positive. We are interested in the case when
$E^{\pm} \gg |\Delta|$, $\delta$ and also exceed the level broadening
$\Gamma_{\alpha}^{i}=\pi\nu |V_{\alpha,i}|^2$
($\nu$ being density of states in the leads) and temperature $T$
(Coulomb blockade region).
In this case we can integrate out
the states with one or three extra electrons. This is equivalent to
Schrieffer-Wolf transformation which is used to obtain the conventional
Kondo model \cite{classics}.
We obtain the following effective low-energy Hamiltonian
\begin{equation}
H_{\rm{eff}}=H^{S=1}+H^{S=1 \leftrightarrow 0}+H_{\rm{eff}}^{\prime}+
H_{\rm{dot}}.
\end{equation}
The first term involves components of the spin-triplet state and resembles
a conventional Kondo Hamiltonian for $S=1$.
\begin{eqnarray}
H^{S=1} & = & \sum_{k k'} \sum_{\alpha \beta=L,R} \sum_{i=1,2} J_{\alpha
\beta}^{(i)}
\left[ \hat{S}_{+} c_{\alpha k' \downarrow}^{(i) \dagger} c_{\beta k
\uparrow}^{(i)}
+\hat{S}_{-} c_{\alpha' k \uparrow}^{(i) \dagger} c_{\beta k
\downarrow}^{(i)}
+\hat{S}_{z} (c_{\alpha' k \uparrow}^{(i) \dagger} c_{\beta k
\uparrow}^{(i)}
             -c_{\alpha' k \downarrow}^{(i) \dagger} c_{\beta k
\downarrow}^{(i)}) \right] \nonumber \\
        & = & \sum_{k k'} \sum_{\alpha \beta=L,R} \sum_{i=1,2} J_{\alpha
\beta}^{(i)}
\Bigl[ \sqrt{2}(f_{11}^{\dagger}f_{10}+f_{10}^{\dagger}f_{1 -1})
                       c_{\alpha k' \downarrow}^{(i) \dagger} c_{\beta k
\uparrow}^{(i)}
+\sqrt{2}(f_{10}^{\dagger}f_{11}+f_{1 -1}^{\dagger}f_{10})
                       c_{\alpha k' \uparrow}^{(i) \dagger} c_{\beta k
\downarrow}^{(i)}  \nonumber \\
& & +(f_{11}^{\dagger}f_{11}-f_{1 -1}^{\dagger}f_{1 -1})
                      (c_{\alpha k' \uparrow}^{(i) \dagger} c_{\beta k
\uparrow}^{(i)}
                      -c_{\alpha k' \downarrow}^{(i) \dagger} c_{\beta k
\downarrow}^{(i)}) \Bigr].
\end{eqnarray}
Here we have introduced pseudo-fermion operators $f_{SM}^{\dagger}$
($f_{SM}$) which
create (annihilate) the state $|SM \rangle$. It is required that
$\sum_{SM} f_{SM}^{\dagger} f_{SM} =1$.
The second term in $H_{\rm{eff}}$ describes the conversion between
the spin-triplet and singlet states accompanied by interchannel scattering of
conduction electrons
\begin{eqnarray}
H^{S=1 \leftrightarrow 0} =
\sum_{k k'} \sum_{\alpha \beta=L,R} \Bigl\{ \tilde{J}_{\alpha \beta}
\Bigl[\sqrt{2}(f_{11}^{\dagger}f_{00}-f_{00}^{\dagger}f_{1 -1})
                       c_{\alpha k' \downarrow}^{(1) \dagger} c_{\beta k
\uparrow}^{(2)}
+\sqrt{2}(f_{00}^{\dagger}f_{11}-f_{1 -1}^{\dagger}f_{00})
                       c_{\alpha k' \uparrow}^{(1) \dagger} c_{\beta k
\downarrow}^{(2)}   \nonumber \\
-(f_{10}^{\dagger}f_{00}+f_{00}^{\dagger}f_{10})
                      (c_{\alpha k' \uparrow}^{(1) \dagger} c_{\beta k
\uparrow}^{(2)}
                      -c_{\alpha k' \downarrow}^{(1) \dagger} c_{\beta k
\downarrow}^{(2)}) \Bigr]
+\tilde{J}_{\alpha \beta}^{*} [1 \leftrightarrow 2] \Bigr\}.
\end{eqnarray}
The third term $H_{\rm{eff}}^{\prime}$ represents the scattering processes
without change of the
dot state and is not relevant for  the current discussion.
The coupling constants are given by
$
J_{\alpha \beta}^{(i)}  =    V_{\alpha,i}V_{\beta,i}^{*}/(2 E_c),
\tilde{J}_{\alpha \beta}  =  V_{\alpha,1}V_{\beta,2}^{*}/(2 E_c),
$
where $1/E_c=1/E^+ +1/E^-$.
The Hamiltonian of the dot reads
\begin{equation}
H_{\rm{dot}}=\sum_{S,M} E_S f_{SM}^{\dagger}f_{SM}.
\end{equation}

To avoid the complication due to the fact that there are two leads
$\alpha=L,R$, we perform a unitary
transformation for electron modes in the leads along the lines of 
Ref.\ \cite{Glazman};
$c_{k \sigma}^{(i)}=(V_{L,i}c_{L,k \sigma}+V_{R,i}c_{R,k \sigma})/V_i$,
$\bar{c}_{k \sigma}^{(i)}=(V_{R,i}^*c_{L,k \sigma}-V_{L,i}^*c_{R,k
\sigma})/V_i$,
with $V_i=\sqrt{|V_{L,i}|^2+|V_{R,i}|^2}$. The modes $\bar{c}_{k \sigma}^{(i)}$
are not coupled to the quantum dot and shall be disregarded.
The coupling constants for modes  $c_{k \sigma}^{(i)}$
become
\begin{eqnarray}
J^{(i)} & = & \frac{|V_{L i}|^2+|V_{R i}|^2}{2E_c}, \\
\tilde{J} & = & \frac{1}{2E_c}\frac{(V_{L 1}^2+V_{R 1}^2)(V_{L 2}^2+V_{R
2}^2)}
{\sqrt{(|V_{L 1}|^2+|V_{R 1}|^2)(|V_{L 2}|^2+|V_{R 2}|^2)}}.
\end{eqnarray}
The spin-flip processes included in our model are shown in the inset of
Fig.~1(b).

We calculate the Kondo temperature $T_K$
with the poor man's scaling technique \cite{Anderson,multiK}.
By this method, we can properly consider the energies from $T_K$ to
the upper cutoff. We concentrate on evaluating the exponential part
of $T_K$.
We assume constant density of states in the leads $\nu$ in the
energy band of
$[-D, D]$. By changing the energy scale from $D$ to $D-|d D|$, we obtain a
closed form of
the scaling equations for $J^{(1)}$, $J^{(2)}$, and $\tilde{J}$ in two limits.

In the first limit, the energy difference $|\Delta|$ is negligible 
($|\Delta| \ll  D$) and $H_{\rm{dot}}$ can be safely disregarded.
 The scaling equations are best presented in the following matrix form:
\begin{equation}
\frac{d}{d\ln D}
           \left( \begin{array}{cc}
           J^{(1)} & \tilde{J} \\
           \tilde{J}^{*} & J^{(2)} \end{array} \right)
=-2 \nu
           \left( \begin{array}{cc}
           J^{(1)} & \tilde{J} \\
           \tilde{J}^{*} & J^{(2)} \end{array} \right)^2.
\label{eq:scalA}
\end{equation}
The equations can be readily rewritten for eigenvalues of the matrix,
$J_{\pm}=(J^{(1)}+J^{(2)})/2 \pm
\sqrt{(J^{(1)}-J^{(2)})^2/4+|\tilde{J}|^2}$.
The larger one, $J_{+}$, diverges faster upon decreasing the bandwidth $D$
and hence
determines $T_K$. 
If the equations remain valid till the scaling ends ($|\Delta| \ll T_K$),
the Kondo temperature is 
$T_K(0)=D_0 \exp [-1/2\nu J_+]$.  Here 
$D_0$ is the initial bandwidth
given by $\sqrt{E^+ E^-}$ \cite{Haldane}.
In another limiting case, $\Delta \gg D$. In this case
the ground state of the dot is spin triplet and the singlet state can be
disregarded.
$J^{(1)}$ and $J^{(2)}$ evolve independently 
\begin{equation}
\frac{d}{d\ln D} J^{(i)} = -2\nu J^{(i) 2},
\label{eq:scalB}
\end{equation}
and $\tilde J$ does not change.
If these equations 
remain valid in the whole scaling region ($\Delta > D_0$),
it yields $T_K(\infty) =D_0 \exp [-1/2\nu J^{(1)}]$.
Here we assume $J^{(1)} \ge J^{(2)}$. 
This is the Kondo temperature for spin-triplet localized spins \cite{Okada}.

To determine $T_K$ in the intermediate region,
$T_K(0) \ll \Delta \ll D_0$, we match
the solutions of Eqs.\ (\ref{eq:scalA}) and (\ref{eq:scalB})
at $D \simeq \Delta$. $\tilde J$ saturates at this point 
while $J^{(1)}$ and $J^{(2)}$ continue to grow with decreasing $D$. 
This yields power law dependence on $\Delta$
\begin{equation}
T_K(\Delta)=T_K(0)\cdot \left( T_K(0)/\Delta \right)^{\gamma},
\label{eq:TK}
\end{equation}
where $
\sqrt{\gamma}=|\tilde{J}|/[\sqrt{ (J^{(1)}-J^{(2)})^2/4+|\tilde{J}|^2}+
|J^{(1)}-J^{(2)}|/2]$. The exponent $\gamma$ appears to be nonuniversal,
depending on a ratio of the initial coupling constants.
In general, $0 <\gamma \le 1$. 
For $\Delta<0$, all the coupling constants saturate 
and no Kondo effect is expected,
provided $|\Delta| \gg T_K(0)$.

In a simple case of the identical couplings, $J^{(1)}=J^{(2)}=\tilde{J}$
$(\equiv J)$, $T_K(0)=D_0 \exp [-1/4\nu J]$.
For $\Delta>0$, $T_K$ decreases with increasing $\Delta$ as
$T_K(\Delta)=T_K(0)^2/\Delta$
($\gamma=1$ in Eq.\ (\ref{eq:TK})) and finally converges to
$D_0 \exp [-1/2\nu J]=T_K(0)^2/D_0$. For $\Delta<0$, $T_K$ drops to zero
suddenly at $|\Delta| \sim T_K(0)$.
The dependence of the Kondo temperature on $\Delta$ is schematically shown in
Fig.~1(b).

We have discussed so far the case of $C_1=C_2=1$ 
in Eq.\ (\ref{eq:singlet}) for the
spin-singlet state.
This is not required by symmetry and $C_1 \ne C_2$ in general.
To justify the assumption we made, let us consider the renormalization
equations for $C_1 \ne C_2$.
The coupling constants
$\tilde{J}_1=C_1 \tilde{J}$ and $\tilde{J}_2=C_2 \tilde{J}$ are
renormalized now in a different way, involving
the scattering processes without
spin flip in the dot
\begin{equation}
H_{\rm{eff}}^{\prime}=\sum_{k k' \sigma}\sum_{i=1,2}
        \left[ J^{\prime (i)} c_{k' \sigma}^{(i) \dagger} c_{k \sigma}^{(i)}
\sum_M f_{1 M}^{\dagger}f_{1 M}
              + J^{\prime\prime (i)} c_{k' \sigma}^{(i) \dagger} c_{k
\sigma}^{(i)} f_{00}^{\dagger}f_{00}
        \right].
\end{equation}
General scaling equations for $|\Delta| \ll D$ are given by
\begin{eqnarray}
d \tilde{J}_1/ d \ln D & = & -2\nu (J^{(1)}+J^{(2)})\tilde{J}_1 + \nu J'
\tilde{J}_1
\label{eq:scalC} \\
d \tilde{J}_2/ d \ln D & = & -2\nu (J^{(1)}+J^{(2)})\tilde{J}_2 - \nu J'
\tilde{J}_2 \\
d J'/ d \ln D & = & 8 \nu ( |\tilde{J}_1|^2 - |\tilde{J}_2|^2 ) \\
d J^{(i)} / d \ln D & = & -2\nu \left[ J^{(i) 2} + (|\tilde{J}_1|^2 +
|\tilde{J}_2|^2 )/2 \right]
\label{eq:scalD}
\end{eqnarray}
where $J'=J^{\prime (1)}-J^{\prime (2)}-J^{\prime\prime (1)}+J^{\prime\prime
(2)}$. When $\Delta \gg D$,
the equations are identical to Eq.\ (\ref{eq:scalB}).
Our point is that
if we concentrate on the most rapidly divergent
solutions of Eqs.\ (\ref{eq:scalC})-(\ref{eq:scalD}), which are proportional
to $1/\ln D$, $\tilde{J}_1$ and $\tilde{J}_2$
appear to be the same.
To this leading order, the renormalization is the same as given by 
Eq.\ (\ref{eq:scalA}).
Consequently the Kondo temperature is the same as that in the case of
$C_1=C_2=1$, apart from a prefactor.

We calculate perturbation corrections to conductance when
$T_K \ll T \ll E_c$.
The third order perturbations in $J$'s yield 
the logarithmic corrections $G_K$
typical for the Kondo
effect \cite{classics}. At  $T \gg |\Delta|$,
\begin{eqnarray}
G_K/(2e^2/h)= \sum_{i=1,2}
\frac{4\Gamma_L^{i}\Gamma_R^{i}}{(\Gamma_L^{i}+\Gamma_R^{i})^2}
6\pi^2 J^{(i)}\nu \left[
(J^{(i)}\nu)^2+|\tilde{J}\nu|^2\right]
\ln \frac{D_0}{T} \nonumber \\
+\frac{2(\Gamma_L^{1}\Gamma_R^{2}+\Gamma_L^{2}\Gamma_R^{1})}
{(\Gamma_L^{1}+\Gamma_R^{1})(\Gamma_L^{2}+\Gamma_R^{2})}
12\pi^2 \left[J^{(1)}\nu+J^{(2)}\nu \right]
|\tilde{J}\nu|^2 \ln \frac{D_0}{T}.
\end{eqnarray}
The interplay between
the spin-singlet and triplet states largely enhances the conductance.
In the opposite case, $T \ll |\Delta|$, the interplay becomes  
less effective and the logarithmic corrections become smaller
\begin{equation}
G_K/(2e^2/h)=\sum_{i=1,2}
\frac{4\Gamma_L^{i}\Gamma_R^{i}}{(\Gamma_L^{i}+\Gamma_R^{i})^2}
8\pi^2 J^{(i)}\nu \left[
(J^{(i)}\nu)^2 \ln \frac{D_0}{T}
+|\tilde{J}\nu|^2 \ln \frac{D_0}{\Delta} \right]
\end{equation}
for $\Delta>0$ and
disappear
for $\Delta<0$. If one varies $\Delta$ at fixed $T$, 
one sees the enhanced conductance for $|\Delta| \ll T$.

We believe that the conductance approaches the unitary limit $2e^2/h$ at
$T \ll T_K$ \cite{Glazman} in our model. However, not all the terms
in $G_K$ can be renormalized to a universal function $G(T/T_K)$.
Because of the multichannel nature of our model, one should expect
nonuniversal logarithmic terms along with the universal ones.
Those, however, will be much smaller than $2e^2/h$.


Recent experiment \cite{exp} has shown a significant
enhancement of conductance to the values of the order
of $e^2/h$, around the crossing point between spin-singlet 
and triplet states, in vertical quantum dots.
The conductance remains low in the stability domains
of singlet or triplet states. Our results provide
a possible theoretical explanation for the results. The conductance
increase can be attributed to the Kondo effect enhanced
by the competition between the two states,
near the crossing point. The Kondo temperature
elsewhere is probably low in comparison with the actual temperature,
so that no conductance increase is seen. 

In conclusion, we have shown that the competition between spin-singlet
and triplet states enhances the Kondo effect. This may explain recent
experimental observations. We have predicted power law dependence of the Kondo
temperature on the energy difference between the states.

The authors are indebted to L.\ P.\ Kouwenhoven for suggesting the topic
of the research presented, S.\ De Franceschi, J.\ M.\
Elzerman, K.\ Maijala, S.\ Sasaki,
W.\ G.\ van der Wiel, Y.\ Tokura, L.\ I.\ Glazman, M.\ Pustilnik,
and G.\ E.\ W.\ Bauer for
valuable discussions.
The authors acknowledge financial support from the
``Netherlandse Organisatie voor
Wetenschappelijk Onderzoek'' (NWO). M.\ E.\ is also grateful for financial
support from the
Japan Society for the Promotion of Science for his stay at Delft University
of Technology.


  \begin{figure}[hbt]
  \centering
  \vspace*{1cm}
  \epsfig{file=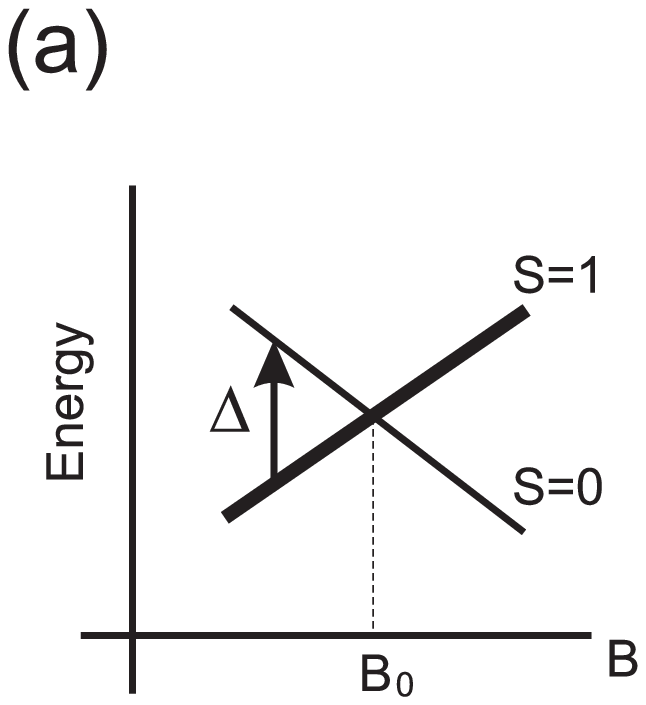,width=1.7in} \hspace{.5cm}
  \epsfig{file=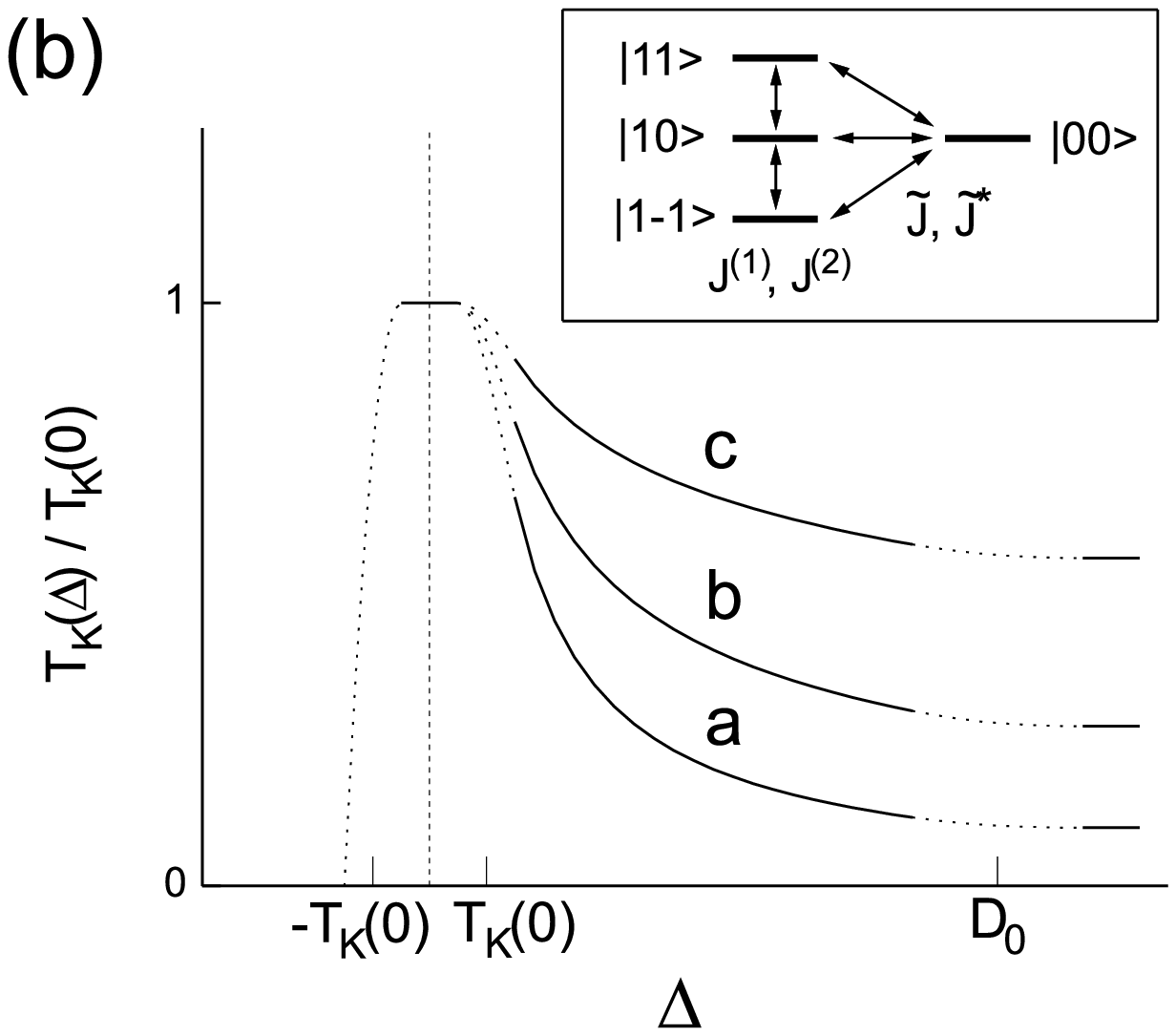,width=3in}
  \vspace*{1cm}
  \caption{(a) The energies of spin-singlet and triplet states in
a quantum dot, as functions of magnetic field $B$.
The energy difference, $\Delta=E_{S=0}-E_{S=1}$, can be controlled by
changing $B$.
$\Delta=0$ at $B=B_0$, where the transition of the ground state occurs.
(b) Schematic drawing of the Kondo temperature, $T_K$, as a function of
$\Delta$. When $\Delta>0$,
$T_K(\Delta)/T_K(0)=(T_K(0)/\Delta)^{\gamma}$ where
{\bf a} $\gamma=1$, {\bf b} 0.5 and, {\bf c} 0.25 at 
$T_K(0) \ll \Delta \ll D_0$ (bandwidth), and
$T_K(\Delta)$ is a constant at $\Delta \gg D_0$. When $\Delta<0$,
$T_K(\Delta)$ drops
to zero suddenly at $|\Delta| \sim T_K(0)$.
Inset: Spin-flip processes in our model. The exchange
couplings $J^{(i)}$ involving spin-triplet states only are accompanied 
by scattering of conduction electrons of channel $i$.
Those involving spin-triplet and singlet states ($\tilde{J}$, $\tilde{J}^*$)
are accompanied by interchannel scattering of conduction electrons.
}
  \end{figure}


\begin{references}
\bibitem{symmetry}
Our main results, enhancement of $T_K$ and its power law dependence
on $\Delta$, are valid also for a simpler model where the
levels belong to the same symmetry representation. In this case, however,
the level repulsion  makes it difficult 
to organize the degeneracy of spin-singlet and triplet states. This is why we
prefer the model in use. Besides, it is experimentally relevant \cite{Tarucha}.
\bibitem{com1}
Considering the matrix element of the Coulomb interaction between
$d_{1 \uparrow}^{\dagger} d_{1 \downarrow}^{\dagger} |0 \rangle$
and $d_{2 \uparrow}^{\dagger} d_{2 \downarrow}^{\dagger} |0 \rangle$,
$\langle 22|e^2/r|11 \rangle$,
one observes that this is indeed the case for $\delta =0$.
\end{references}
\end{document}